# Latent Dirichlet Allocation (LDA) for Topic Modeling of the CFPB Consumer Complaints

Kaveh Bastani[1],*, Hamed Namavari[1,2], Jeffry Shaffer[1,3]

[1]Unifund CCR, LLC, Cincinnati, OH
[2]Economics, College of Business, University of Cincinnati, Cincinnati, OH
[3]Business Analytics, College of Business, University of Cincinnati, Cincinnati, OH

**Abstract -** A text mining approach is proposed based on latent Dirichlet allocation (LDA) to analyze the Consumer Financial Protection Bureau (CFPB) consumer complaints. The proposed approach aims to extract latent topics in the CFPB complaint narratives, and explores their associated trends over time. The time trends will then be used to evaluate the effectiveness of the CFPB regulations and expectations on financial institutions in creating a consumer oriented culture that treats consumers fairly and prioritizes consumer protection in their decision making processes. The proposed approach can be easily operationalized as a decision support system to automate detection of emerging topics in consumer complaints. Hence, the technology-human partnership between the proposed approach and the CFPB team could certainly improve consumer protections from unfair, deceptive or abusive practices in the financial markets by providing more efficient and effective investigations of consumer complaint narratives.

## 1. Introduction

In 2011 the CFPB was created by congress to ensure the protection of consumer interests in many financial markets. The CFPB receives and processes consumer complaints pertaining to various financial services including credit cards, mortgages, bank accounts, student loans, consumer loans, credit reports, payday loans, and debt collection. The raw consumer complaints data is updated every night and publicly available to download from the CFPB's website[2]. This database, known as the largest public collection of consumer financial complaints, includes basic information about the complaints such as submission date, consumer's zip code, the company, the product type, the relevant issue, the consumer narratives, and how the company has addressed the complaint. Making the data publicly available not only allows the financial institutions and their consumers to view the overall quality of their financial products and services, but also encourages the public users such as analysts, data scientists and academicians to explore these information, and build on valuable insights accordingly. Nonetheless, there have been a very few published works documenting formal analysis of CFPB database. In the following, a review of these studies is provided.

---

* Corresponding author, kaveh@vt.edu, 10625 Techwoods Circle, Cincinnati, OH 45242
[2] http://www.consumerfinance.gov/data-research/consumer-complaints/

Ayres et al. (2013) investigate an early analysis of the CFPB consumer complaints at company-specific level as well as zip code demographics level. Their company-specific results demonstrate that companies such as Bank of America, Citibank, and PNC Bank were significantly (at 10% significance level) less timely in responding to complaints than the average financial institutions. Additionally, their results provide a list of companies including Wells Fargo, American Express, and Bank of America whose consumers are significantly more likely than the average consumers to dispute the company's response. Using the zip code information associated with the consumer complaints and the census zip code tabulation areas (ZCTA), they were able to approximate the demographics information with complaints. Subsequently, their regression analysis on the zip code demographics information indicates that the mortgage complaints significantly increased in zip codes with larger proportions of certain populations such as senior citizens, and high school and college graduates.

Littwin (2015) takes a different perspective from Ayres et al. (2013) in analyzing the CFPB database at the government agencies level, i.e., the author explores the reasons why government agencies should process the CFPB consumer complaints and whether these reasons justify the resources that complaint processing requires. She investigates three reasons to process the CFPB consumer complaints, namely to settle consumer disputes, to inform the government agency's regulatory activities, and to generate good will for the government agency. Her regression based analysis demonstrates that (1) the CFPB has been successful in providing a dispute resolution forum for the consumers, (2) the CFPB is strong on regulatory activities and displays a significant commitment in ensuring the consumers complaints are handled by the relevant companies, and (3) there is good will associated with the CFPB activities.

Although Ayres et al. (2013) and Littwin (2015) lead to interesting results from consumer complaints data that should be of interest to the CFPB, financial companies, and their consumers, they both lack an analysis of the consumer complaint narratives that certainly delivers more valuable resources to the financial community. In fact, the narratives provide context to complaints upon which more interesting insights could be built. This requires scholars to review the CFPB complaint narratives in order to extract insights from the documents. The amount of narratives received by the CFPB has been increasing over time. With an increasing rate of the CFPB complaint narratives, there is a need for efficient procedures to review and analyze the narratives. Obviously, manual review of large volume of documents by humans is not feasible as it is a time-consuming process and often subject to human biases.

One trivial solution to the above challenge is labeling the narratives based upon their context at the time of submission. The CFPB has designed a drop-down menu at the CFPB complaint submission portal which allows consumers to choose what issue (e.g. Bankruptcy, Billing disputes, etc.) their narratives are about. These issues are predetermined labels defined by the CFPB experts which describes the primary reason for the consumer complaints. This is the CFPB convention in labeling the consumer complaints. However, this CFPB convention appears to have two shortcomings. First, the consumer selects the label best describing her complaint issue from the drop-down menu. Due to the consumer's lack of knowledge or understanding of the labels, she might select a wrong label (mislabeling), or the issue she would like to complain about might not be described by the labels provided in the drop-down menu. As such, she would be forced to select a label that does not best describe her issue. Second, the consumer complaint might be about multiple issues (labels), but due to the limitation of CFPB labeling convention, he can only select one label.

The above shortcomings could be addressed by a thorough analysis of the narrative data using text mining approaches. Text mining refers to the process of extracting non-trivial patterns or knowledge from unstructured text documents (Aggarwal and Zhai, 2012). Text mining is a much more complex task than data mining (for example the analyses reported in Ayres et al. (2013) and Littwin (2015)) as it involves handling text data that is inherently unstructured and noisy, especially in the realm of consumer complaints.

This paper aims to demonstrate the utility of text mining approaches in analyzing textual data contained in the CFPB consumer complaint narratives. To the best of authors' knowledge, there has been no published paper analyzing the CFPB complaint narratives by text mining approaches. However there have been previous research works exploring the use of text mining techniques in decoding consumer complaints related to other fields such as the National Highway Traffic Safety Administration's (NHTSA) (Ghazizadeh et al., 2014), virtual travel agencies (Lehto et al., 2007), low-cost airlines or low-cost carriers (LCCs) (Yee and Pei, 2014), the hotel industry (Berezina et al., 2016), and e-auction stores (Chen, 2011). These studies utilize semi-automated, or fully automated approaches on the textual narrative data to extract and identify coherent insights/topics in their fields.

The most prominent of these approaches is latent semantic analysis (LSA) which is an efficient approach in identifying/clustering similar narratives by reducing the high dimensionality of the textual data (Deerwester et al., 1990). However, LSA per se is not equipped with any generative probabilistic model, and therefore no attempt can be made to fit a model to the textual data using maximum likelihood or Bayesian estimation. Hence, no topic modeling can be implemented by LSA as there is no generative model available to capture topics. To account for topics, LSA is typically applied with clustering techniques (such as K-means (Hartigan and Wong, 1979) or hierarchical clustering (Zhao and Karypis, 2002)) to be able to cluster the complaint narratives into groups (also known as topics in text mining context). However, the underlying assumption of these clustering techniques is that a consumer narrative only belongs to a single topic, while this might not be the case in consumer complaint narratives. It is very likely that a narrative shares multiple topics, i.e., the consumer might be complaining about multiple subjects (please refer to Section 3.2 and Figure 3 for an explanatory example on a narrative sharing multiple topics). Therefore, there is a need for a text mining approach that overcomes this issue by assigning multiple topics to consumer narratives.

This paper proposes utilizes a probabilistic topic modeling approach known as latent Dirichlet allocation (LDA) (Blei et al., 2003) for the CFPB consumer complaint narratives. The LDA approach extracts the latent topics from the consumer complaint narratives, and simultaneously assigns a probabilistic mixture of these topics into the consumer complaints. The LDA approach does not require any prior annotations about the consumer narratives; instead the topics are learned from the analysis of the original texts. LDA based topic modeling has been applied to different kinds of text data including email (Blei and Lafferty, 2009), scientific abstracts (Griffiths and Steyvers, 2004), and newspaper archives (Wei and Croft, 2006). Interesting applications of LDA have also been reported as a powerful technique to create knowledge and discover useful structure in a stream of literature (Asgari and Bastani, 2017).

To our best knowledge, this is the first study exploring the utility of LDA on consumer complaint narratives provided by the CFPB datasets. The LDA approach enables us to classify unstructured consumer complaints into a mixture of topics through which their underlying content would be revealed. This task would not be possible by human annotations as it requires reading the full text of the narratives, and identifying the similarities between them, which is very difficult and time consuming. These topics are learned from the analysis of the complaint narratives with no need for input from the consumers on the issue. Therefore, the existing challenges with the CFPB labeling convention can be addressed effectively. Furthermore, it would be possible to reveal the trends of these topics over time, and evaluate the effectiveness of the CFPB regulations and financial companies in addressing these topics.For example, if it is observed that a topic trend is decreasing over time it could indicate that the CFPB regulations have taken the relevant topic into account, and the companies have been able to improve their customer service with respect to that topic. In the following, the LDA based appoach is proposed for topic modeling of CFPB consumer complaints, and the results of this analysis are presented and discussed subsequently.

## 2. LDA based Topic Modeling to the CFPB Complaint Narratives

### 2.1 Preparing Consumer Complaint Narrative Data

The CFPB consumer complaint database is publicly available and can be downloaded through the CFPB website[3]. This Database is a collection of 615,273 complaint records including certain information for each complaint such as the date of submission, the product and the issue the complaint is about, the company the complaint was sent to for response, etc. Additionally there is a consumer complaint narrative filed in the database including the consumer descriptions of what happened. However, this field is not populated for all the records as it is available only if the consumer opts to share. This paper focuses exclusively on consumer complaint narratives. Hence, after removing the records with no consumer complaint narratives; the total of 88,295 records remained. It was also observed that from the remaining records, 1492 narratives were repetition of others, hence the repetitive records were excluded, to provide the total of 86,803 records for our text mining analysis.

**2.2 Data Preprocessing**

The proposed text mining analysis starts with preprocessing the data obtained from Section 2.1. Before presenting the data preprocessing tasks, the terminologies used throughout the paper are introduced. A *document* which is defined as a sequence of words and punctuation, refers to any consumer complaint narrative from the CFPB dataset. In our text mining analysis, there are 86,803 documents consistent with the number of records available in the consumer complaint narrative data. A *term* is defined as a word, however term and word are interchangeably used throughout this paper. The *term-document matrix* is a matrix projecting the terms found among all the 86,803 documents into the individual documents. Each row of the term-document matrix corresponds to a term, and each column corresponds to a document. The values in the matrix cells presents the frequency of each term in each document.

Our data preprocessing includes five tasks (see Figure 1 for the summary of these tasks), namely, (1) convert to lowercase, (2) remove special characters and tokenize them into terms, (3) remove stop words, (4) stemming, and (5) construct term-document matrix. The details of each of these tasks is explained in the following. In Step 1 all the documents are converted into lowercase to provide a consistent format for all the documents. Step 2 removes the special characters including punctuation marks (e.g. !%$#&*?,/.;"\) and numbers, as they never contribute to our text mining analysis, and then tokenizes the document into terms. Step 3 is the major task in our data preprocessing. Stop words are generally a set of commonly used words in any language (here English) that should be removed from the document, in order to make the focus on the important words instead. In this analysis two types of stop words, namely, generic stop words (some common words such as "a", "an", "is", etc), and domain specific stop words, are defined.

---

[3] http://www.consumerfinance.gov/data-research/consumer-complaints/#download-the-data.

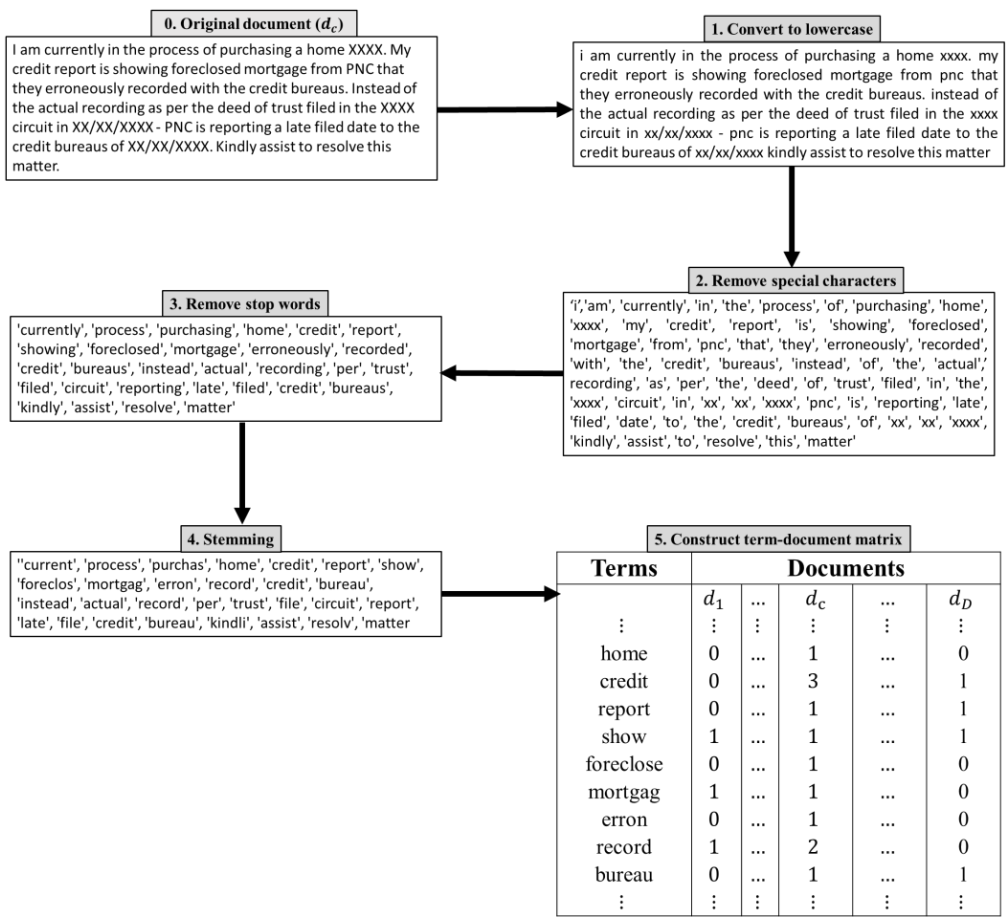

Figure 1. Data preprocessing tasks for a consumer complaint narrative with complaint ID 1438325 (denoted by document $d_c$). The construct term-document matrix task in this figure is a snapshot from the whole matrix including the frequency of some of the terms in document $d_c$.

There is no specific rule in identifying the domain specific words, and it requires a thorough consideration of the objectives of our text mining analysis as well as the domain knowledge. The aim of our analysis is to extract the latent topics from the consumer narratives, hence it would make sense to exclude the companies' names from the documents. This is so, as it is observed that the names of the companies appear so frequently in the full text of the consumer complaints, and that those names essentially do not convey any meaningful message for our topic modeling. Hence, the companies' names (e.g. Bank of America, Wells Fargo, Citibank) were included in the stop words set. Similar to the companies' names, it was found that state names do not contribute to the topic modeling as well and therefore they were also included in the stop words set. More importantly, the CFPB applies masking to the consumer complaint narratives to protect consumers' personal identifiable information or personal sensitive information. As a result, in the text of the consumer narratives, the words "xx""xxxx" and "xxxxxxxx" appear too frequently for the masking purposes, thus, the stop words set also include these words.

The goal of stemming is to reduce the variation in the text data by converting words to their common base form/word stem. For example, "finance", "financial", "finances" and "financing" were converted to "financ". The stemming step is very common in text mining analysis as it helps concentrate the analysis on the base form of the words, rather than differentiating between variations of the words that might cause confusion to the text mining algorithms. The last task in data preprocessing is constructing the term-document matrix. This matrix presents the distribution/frequency of terms (rows) within documents (columns). Indeed, the matrix reduces documents (narratives) into vectors of unique words that are presented as the columns. The term-document matrix is mainly used as an input to most of the text mining algorithms, including LSA, and LDA.

In the case of high dimensionality of the term-document matrix, some dimensionality reduction techniques, most notably LSA can be applied. LSA uses a singular value decomposition (SVD) of the matrix to identify a linear subspace of the original feature space formed by the term-document matrix to capture most of the variance in the document collections. However, LSA per se cannot be applied for the purpose of topic modeling as it is not equipped with any generative probabilistic model. A significant improvement in this regard was initiated from Hofmann (1999) by developing probabilistic LSA model (pLSA). pLSA assigns a probabilistic mixture model to the words in a document, where the mixture components are viewed as representation of topics. Although pLSA is capable of assigning multiple topics to a document, it cannot be used to infer (predict) topic assignments of unseen documents due to lack of a well-defined generative model (Blei et al., 2003).

The aim of this paper is to provide an explanatory analysis of the extracted topics in collection of consumer narratives. However being capable of generating a predictive model to reveal the topic representation of the future complaints makes the text mining analysis more powerful. Hence, pLSA might not be an appealing approach for text analysis of the consumer complaints due to the above shortcoming. Fortunately, LDA overcomes this issue by treating the topics as latent random variables estimated from the textual data using hierarchical Bayesian analysis. As a result it can be applied to both topic modeling and predictive modeling of the CFPB consumer complaint narratives. The general overview of the LDA approach is presented in the following Section.

## 2.3 Latent Dirichlet Allocation (LDA)

LDA was first developed by Blei et al. (2003) as a generative probabilistic modeling approach to reveal hidden semantic structures in a collection of textual documents. The basic idea is that each document exhibits a mixture of latent topics wherein each topic is characterized by a distribution over the words (unique words in the collection of documents), and the relative importance of the topics — captured in the form of different weights — vary from document to document. The underlying generative process of LDA is illustrated in Figure 2, and the corresponding variables are pictorially explained in Figure 3.

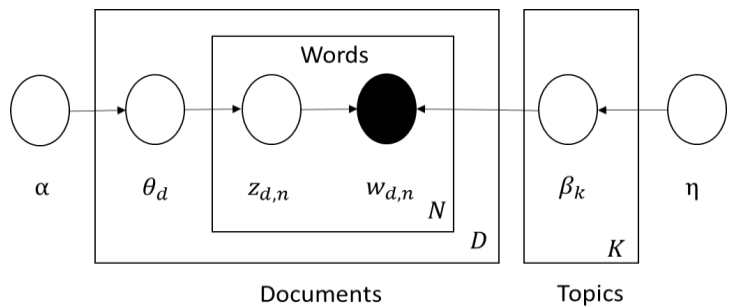

Figure 2. Latent Dirichlet allocation (LDA) for topic modeling

As shown in the diagram, the words within documents $w_{d,n}$ $\forall n = 1, ..., N, \forall d = 1, ..., D$ are observable variables (any words in the collection of complaint narratives in Figure 3), while the other components, consisting of the topics $\beta_k \, \forall k = 1, \ldots, K$ (the distributions over words at top right of Figure 3), the topic distribution per document $\theta_d \, \forall d = 1, \ldots, D$ (the histogram at the bottom right of Figure 3), and the per-word topic assignment $z_{1:D,1:N}$ (the colored words in Figure 2), are not known. These latter items represent unobservable/hidden variables (white circles in Figure 2), that should be estimated from the analysis of observable variables (shaded circles in Figure 2). Parameters $\eta$ and $\alpha$ are the hyperparameters for prior distributions of $\beta_k$ and $\theta_d$, respectively. As such, the values larger than one for these hyperparameters lead to smooth distributions over topics or words, while as values below one result in sparse distributions over a fewer topics or words. The boxes represent the plate notations used to illustrate the replications, i.e., the $K$ plate denotes the number of topics, the $N$ plate denotes the total number of unique words within documents, and the $D$ plate denotes the number of documents.

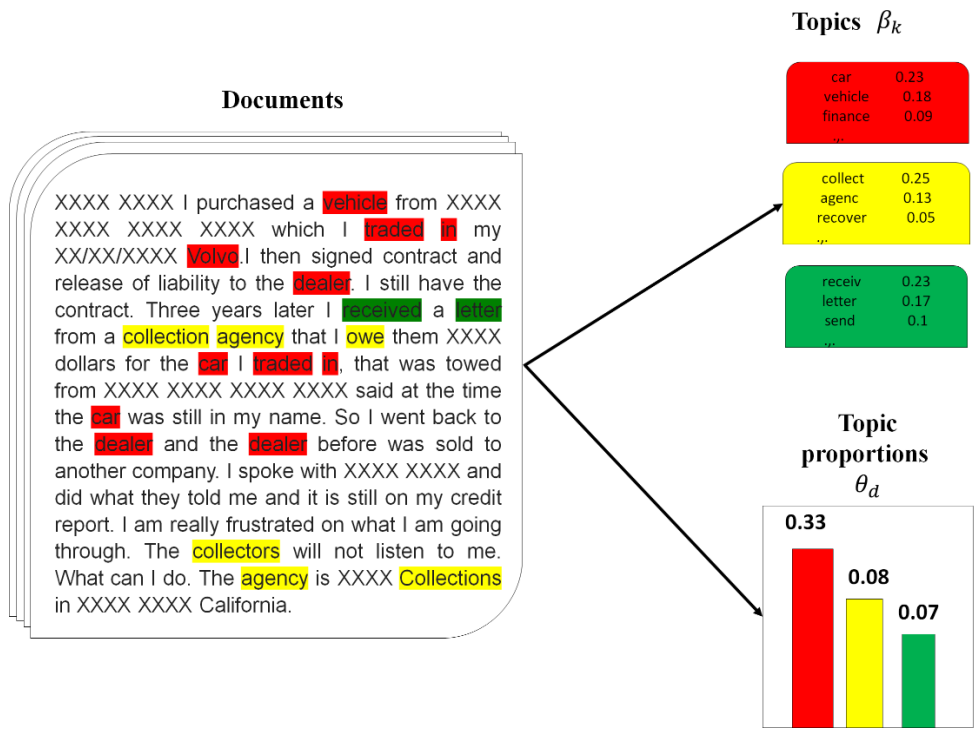

Figure 3. Illustrative example of LDA components and outputs on a complaint narrative with complaint ID 1313544. In this example, $w\,d\,,n$ denotes any words in the complaint narrative, e.g. "car", and $z_{d,n}$ shows per-word topic assignment, e.g. the word "car" is assigned to topic 1 (red color), hence, $z_{d,n} = 1$. The complaint narrative is summarized as a mixture of top three topics $\beta_1$, $\beta_2$, and $\beta_3$ with the highest weights (proportions) denoted by $\theta_d$ and shown in the histogram at the bottom right of the figure. Each topic is a distribution over the words; e.g. $\beta_1$ is a distribution over "car", "vehicle" and "finance" with the probability of 0.23, 0.18, and 0.09 (only top three words with the highest probabilities are shown due to space limit). Labels of the topics are as follows: Auto loan/dealership, collection agency, and communication.

The arrows indicate conditional dependencies among the variables in the following fashion: per-word topic assignment $z_{1:D,1:N}$ is dependent on the topic distribution per document $\theta_d$, and the observed word in each document $w_{d,n}$ is dependent on $z_{1:D,1:N}$ and all the topics $\beta_k$'s. The conditional dependencies allow us to define the joint distribution of observed and unobserved variables. Consequently, LDA has the notable benefit of utilizing Bayesian learning to infer unobservable constructs by computing their posterior distribution from the joint distribution. Learning the unobservable variables enables us to elucidate the semantic structure hidden in the documents. More specifically, LDA provides two main outputs, namely, topics $\beta_k$ and their weights (importance) in each document $\theta_d$. Figure 3 illustrates explanatory outputs of LDA on a collection of complaint narratives in the CFPB dataset. Due to space limitations for this illustration, the outputs include only a single narrative, as presented in this figure. The outputs themselves consist of the three topics exhibiting the largest weights, wherein each topic is represented as a combination of top three words with the highest probability of occurrence. However, to deliver the content of the topics more precisely, it would be helpful to label them. In other words, using human judgments and intervention, the topics can be labeled based on the semantic similarities of their major words (Chang et al., 2009). In this correspondence, the labels associated to these three topics are also provided in the figure.

Figure 3 illustrates the use of LDA in exploring a large volume of documents. Here, instead of dealing with whole texts of, which is a time-consuming task, each document is represented as a combination of the topics extracted by LDA. Since the importance of each topic to each document is learned by LDA, it is possible to identify which composition of topics best represents the essence of each document. In other words, each document can be summarized as a combination of topics with the largest weights (e.g. top three topics in Figure 3). This makes LDA an appealing method to automatically review and explore large volume of complaint narratives received by the CFPB and overcome the burdens associated to the labeling convention of the CFPB.

## 3. Topic Modeling for CFPB Consumer Complaints

### 3.1 Topic Modeling Implementations

Both data preprocessing tasks and LDA analysis were conducted in Python. The nltk library (Joakim, 2012) was used to preprocess the complaint narratives following the tasks explained in Section 2.2. The term-document matrix is the outcome of these tasks which is imported into the genism Python library (http://radimrehurek.com/gensim) for conducting LDA analysis. As it can be seen from Figure 2, the LDA algorithm requires inputs for two specific hyperparameters, namely α and η, and the number of topics $K$. The hyperparameters in LDA are both smoothing parameters on the distributions, i.e. α is a topic smoothing parameter, and η is a word/term smoothing parameter. Values larger than one for these parameters lead to more even distributions, while as values below one result in more concentrated distributions over topics or words. As suggested from most of the topic modeling analyses in the literature (Blei et al, 2003; Blei and Lafferty, 2009; Kaplan and Vakili, 2015; Blei, 2012), the value of 0.1 for both hyperparameters results in semantically meaningful topics. This has been also verified through our numerical analysis; it was observed that a default value 0.1 for these hyperparameters α and η lead to more meaningful outcomes.

The LDA algorithm also requires the number of topics $K$ as input. Many of the topic modeling analyses using LDA in literature (Blei et al, 2003; Blei and Lafferty, 2009; Kaplan and Vakili, 2015; Blei, 2012) determine the number of topics by trial and error procedures. They try different values for $K$, and select the value producing more meaningful topics as outcome (the typical number selected is 100). Recently a novel nonparametric Bayesian topic modeling algorithm named as hierarchical Dirichlet processes (HDP) has been developed (Teh et al., 2012), which is capable of determining the number of topics on its own during posterior inferences in Eq. (2). However, the performance of HDP was not as good as LDA in terms of generating semantically meaningful topics. Hence, the LDA algorithm was yet preferred for the CFPB topic modeling in this paper. Following the trial and error procedures, topic number $K = 40$ was selected as it provided more meaningful topics accordingly.

**3.2 Topic Representation and Assignments for CFPB Consumer Complaints**

The outcomes of LDA algorithm are topics $\beta_k$, per-document topic assignments $z_{d,n}$, and topic proportions $\theta_d$. The results of LDA analysis on the CFPB consumer complaints are presented in the following. The topic outcomes are shown in Table 1. As discussed earlier, the topics are distributions over words; the top ten words with the highest probability (most frequency) derived from posterior distribution $\beta_k$ ($\forall k = 1, ..., K$) are provided for each topic in Table 1. However, in order to characterize the underlying content of the topics, it would be easier to label them rather than presenting them as combinations of words. Unfortunately automatic labeling of topics is not possible as discovering the topics is an unsupervised learning process. It rather requires human judgments and intervention to examine the coherence and meaningfulness of the topics, and subsequently label them through their judgment (Chang et al., 2009). As a result, after extracting the topics using LDA, the authors validated and labeled the topics, which can be seen in the third column of Table1.

The next outcomes from LDA are per-document topic assignments $z_{d,n}$, and topic proportions $\theta_d$. In the context of the CFPB, it would be interesting to represent the topics associated with each complaint, along with their proportions (relevance). There were 86,803 consumer complaints analyzed in this paper. Therefore presenting the topic assignments for each of these documents is not practical at all. Figure 3 illustrates one example from consumer complaint narratives in the CFPB dataset, and the outcomes of LDA on topic assignment and proportions. The complaint ID related to this narrative is 1313522, which is about a company called Data Line Credit Corp. The top five topics with highest proportions assigned to this narrative are: Auto Loan/Dealership (0.325), Collection Agency (0.082), Communication (0.073), Credit Reporting (0.063), and Customer Service (0.061).

The relevant words to each of the topics in the narrative are being colored in correspondence with their associated topics. From these words, it can be verified that LDA has successfully determined the relevant topics. The reason is that the words related to the topic of Auto Loan/Dealership (e.g. vehicle, Volvo, trade in, dealer, car) has appeared more in the narrative compared to the words related to the other

Table 1. Extracted topics for the CFPB consumer complaints using LDA

| ID | Topics | Label |
|---|---|---|
| 0 | payment late made due make appli month past day miss | Payment/Late Payment |
| 1 | receiv letter sent send mail state request email document notic | Communication |
| 2 | loan student borrow privat navient repay lender default defer forbear | Loan/Student Loan |
| 3 | car vehicl financ dealership dealer ticket book drive trade truck | Auto Loan/Dealership |
| 4 | servic custom repres manag transfer spoke depart supervisor cancel speak | Customer Service |
| 5 | check cash advanc clear return wrote flagstar seiz present payabl | Check |
| 6 | file complaint cfpb case complain respond clerk district bsi compliant | CFPB |
| 7 | home hous equiti repair inspect buy door damag sell valu | Home Equity |
| 8 | call phone number person stop time answer messag harass work | Harassment |
| 9 | credit report remov bureau show neg correct inform agenc transunion | Credit Reporting |
| 10 | co program school class colleg signer enrol student region educ | Education |
| 11 | account close open activ statu inaccuraci limit creat past restrict | Account Management |
| 12 | purchas express store buy product order return replac item refund | Store Purchases |
| 13 | consum financi protect practic decept abus institut mislead repossess factor | Consumer Protection |
| 14 | bill insur medic polici hospit cover coverag flood doctor health | Medical Debt/Bill |
| 15 | disput item investig equifax verifi delet inform inaccur valid creditor | Dispute |
| 16 | inquiri author unauthor fraudul tx evict code entitl hard measur | Unauthorized Inquiries |
| 17 | famili mother incom daughter member father live parent employ sister | Family |
| 18 | ident theft fraudul polic victim block ftc stolen affidavit inform | Identity Theft |
| 19 | modif mortgag loan foreclosur home document review incom program trial | MLMF[4] |
| 20 | address name number secur social person live verifi inform licens | Address |
| 21 | violat law act feder fair right note legal assign trust | FDCPA[5] |
| 22 | credit score line limit hard maci pull drop impact affect | Credit Score |
| 23 | agreement close contract apprais sign settlement lender cost agre valu | Settlement |
| 24 | bankruptci file judgment court discharg chapter judgement record public includ | Bankruptcy |
| 25 | interest rate princip refin lower high higher increas principl reduc | Interest Rate |
| 26 | card credit use discov debit reward cancel activ visa declin | Credit and Debit Card |
| 27 | mortgag tax escrow properti insur increas refin sold home ir | Escrow Account |
| 28 | debt collect valid owe collector creditor origin alleg ceas settl | Debt Collector/Issue |
| 29 | attorney court legal employ law firm serv garnish lawyer threaten | Attorney/Legal Actions |
| 30 | husband wife ex divorc heloc name primari child certif spous | Divorce and Ex-spouse |
| 31 | fraud claim investig fraudul alert depart crimin commit activ suspect | Fraud alert |
| 32 | charg fee overdraft refund transact revers merchant waiv assess post | Charged/Overdraft fees |
| 33 | sale properti short lien titl releas buyer sell sold seller | Property Sale |
| 34 | fund money deposit transact branch debit withdraw bank transfer hold | Fund and Deposit |
| 35 | collect agenc recoveri portfolio reinsert invoic enhanc solut central erc | Collection Agency |
| 36 | offer promot point requir bonu term benefit mile condit receiv | Rewards and Promotions |
| 37 | system onlin error websit access free site comput log correct | Online Service |
| 38 | paid amount balanc statement owe full show total due payoff | Payoff |
| 39 | applic approv paperwork process appli deni fill delay discrimin denial | Application |

---

[4] Mortgage/Loan Modification and Foreclosure (MLMF)
[5] Fair Debt Collection Practices Act (FDCPA) is a federal law that controls and restricts the activities of third-party debt collectors.

topics. Hence, LDA assigned the highest proportion to this topic, and following the same logic, it assigned the other topic proportions based on the frequency of their relevant words appeared in the narrative. Figure 3 illustrates only one example of the utility of LDA in topic assignments for the CFPB consumer complaints. The authors created a very interesting visualization (viz) which is available on Tableau public. The readers are strongly suggested to check out the viz on:

https://public.tableau.com/profile/hamed#!/vizhome/CFPBTopicModeling/CFPBTopicModeling.

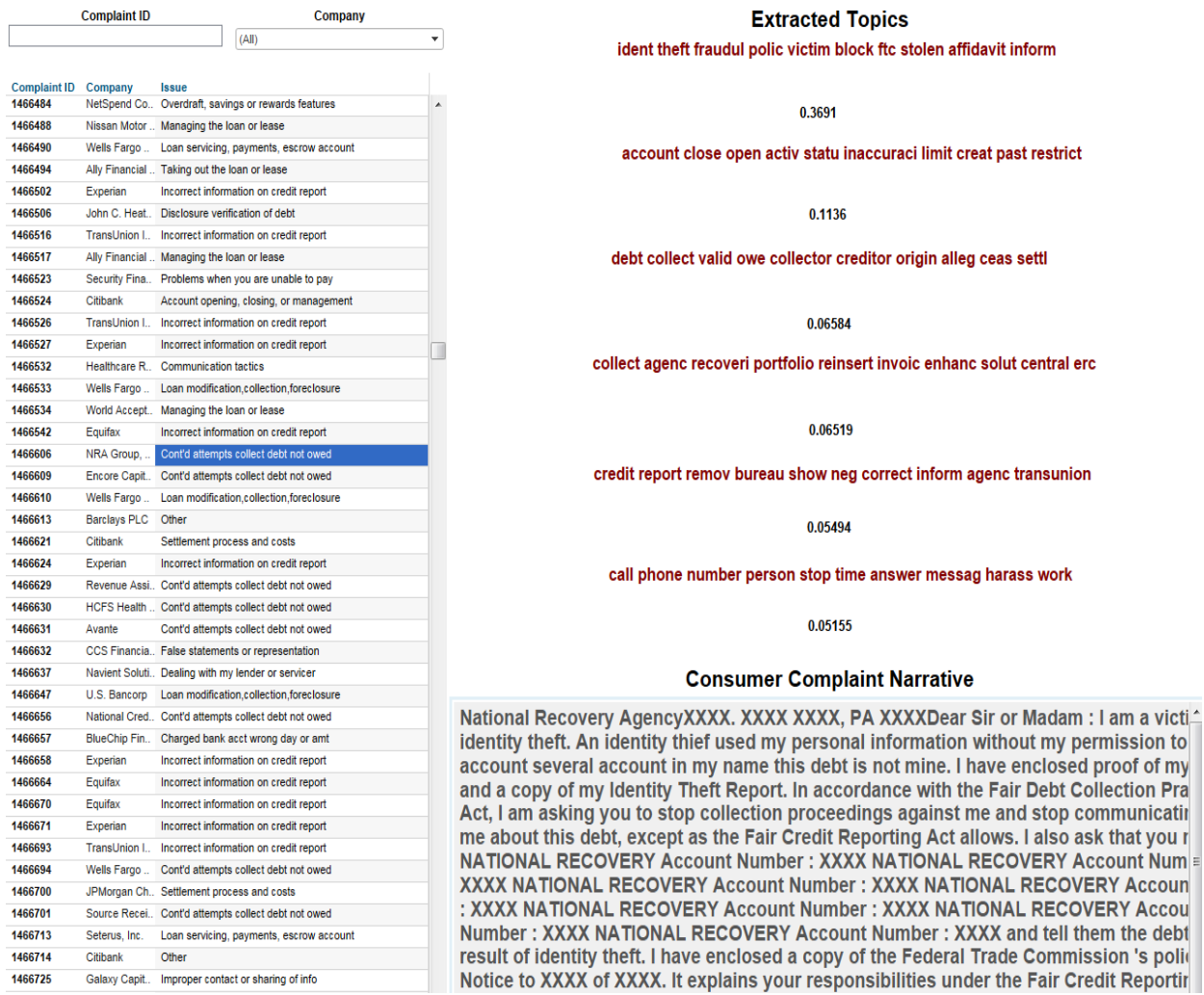

Figure 4. Overview of the Tableau viz on topic assignments and underlying topic proportions for the CFPB consumer complaints.

This viz enables the users to explore all the extracted topics from the CFPB, as well as review the topic assignments for each of the consumer complaints. Furthermore as this is an interactive viz, the users would also be able to select the financial companies they are interested in to evaluate the most frequent topics involved with their consumer complaints. This is a very interesting feature of the viz as through which it would be possible to identify the quality and customer satisfaction issues with these financial companies, and investigate how they have been responsive over time to these issues. For illustration purpose, an overview of the created viz on topic assignments is presented in Figure 4.

On the left of the Figure 4, the complaint ID, company, and the relevant issues of the CFPB complaints are provided. The user is able to select the complaint ID of interest, and review the corresponding complaint narratives, as well as their topic assignments and proportions (top 6 topics are presented with their proportions in the viz). For example, in this viz the complaint ID 1466606 with company NRA Group is selected (the consumer narrative related to this complaint can be found at the bottom of the viz), and the assigned topics with their proportions are shown in the right hand side of the viz.

### 3.3 The CFPB Consumer Complaint Semantic Similarities

Once the topics are extracted from the CFPB, and the topic assignments are carried out using LDA, it would be possible to investigate the semantic similarities of the CFPB consumer complaints through their topic assignments. In other words, the consumer complaints sharing the same topic are supposed to be semantically similar. The authors' Tableau viz also includes the topic similarity analysis. However, since presenting the similarities between all the complaints in the paper is not practical at all, the authors only refer to an example, but strongly suggest the readers to check the Tableau viz.

In this example the topic "Divorce and Ex-spouse" is selected to view the similarities of the CFPB consumer complaints. Note that LDA assigns a mixture of topics with various proportions to a document (consumer narrative), where the "Divorce and Ex- spouse" topic could be the first, second, or even fifth element in that mixture. Therefore, it is expected that the documents contain it as their first or second topic (i.e. with highest or second highest proportions) should be more similar than the documents contain it as their e.g.

fifth topic. Table 2 illustrates two exemplary complaints including "Divorce and Ex-marriage" as their first, third, and fifth topic[6]. The complaint IDs 1856929 and 1872426 both contain "Divorce and Ex-marriage" as their first topic. Reviewing their narratives indicates that the topic relevant words (e.g. ex, divorce, wife, and husband) are very common and frequent in the texts, consequently make these two narratives semantically similar. Indeed both consumers complaint about some financial issues (e.g. debt) owned by their ex-spouse, but instead they have been mistakenly affected.

Table 2 also shows "Divorce and Ex- spouse" appeared as Topic 3 (the third highest proportion) for complaint IDs 1374157 and 1774899. The topic relevant words such as "husband" and "wife" appear not that frequently (only two appearances) in the texts. From their content it can be understood the semantic similarity between these two complaints in terms of "Divorce and Ex-marriage" is low (lower than the previous complaint IDs). But apparently since "husband" and "wife" appeared twice in both of their narratives, LDA assigns "Divorce and Ex-marriage" as the third element in the mixture of topics derived for these documents. Finally, complaint IDs 1302942 and 1291437 were assigned with Divorce and Ex-marriage as their Topic 5. Obviously it is expected that the semantic similarities in terms of this topic should be very low (the lowest in this example).

In general, this analysis verifies that the documents sharing the topics with higher level of proportions, e.g. Topic 1 or 2 are more semantically similar in terms of that topic rather than the documents holding the same topic as their Topic 4 or 5. As mentioned earlier a more comprehensive interactive analysis of the CFPB complaint similarities are provided in the Tableau viz which is available online. An overview of the created viz on document similarities is provided in Figure 5. On the left hand side of the viz the topics are shown. By clicking on each topic, the complaint IDs related to the narratives including the selected topic as their Topic 1 (1st most probable topic), Topic 2 (2nd most probable topic), …, Topic 5 (5th most probable topic) are updated in the viz. For example, in Figure 5, the user was interested in viewing the semantic similarities of the CFPB complaints containing "Identity Theft" as one of their top 5 topics. By hovering over the complaint IDs the user is able to read the raw narratives (e.g. here complaint ID 1408667), and evaluate the semantic similarities of the complaints in terms of "Identity Theft" topic.

---

[6] The complaint narratives provided in this table might contain some typos and misspellings. The reason is that these narratives are taken exactly "as it is" from the CFPB datasets, and no editing was done in order to present the raw data to the readers.

Table 2. Example of the CFPB consumer narrative similarities sharing "Divorce and Ex-spouse" as Topic 1, Topic 3, and Topic 5. The topic relevant words are highlighted in yellow in the narratives

| Topic 1 |
|---|
| Complaint ID (1856929) --- I have XXXX joint mortgages (I accumulated with my ex-wife of almost 20 years ) my credit report and I only own XXXX home. On XXXX XXXX, 2011 I was amicably divorced without legal representation. My divorce decree clearly list XXXX of the mortgages as the debt of my ex-wife. Number XXXX was gifted to my ex-wife and there is a quit claim deed on file. I have no ownership and yet I am being held hostage whenever my credit is reviewed. <br><br> Complaint ID (1872426) --- This a debt that my ex husband has done and am not paying for it .my ex-husband has used my name to get a credit card in my name without my permission and am paying for a debit that is not mine we have been divorced since 2001 and I have no contact with him and I am sick and tried of getting this bills that are not mine I don't even have a credit card because of him. |
| **Topic 3** |
| Complaint ID (1374157) --- I have a balance with Paypal Credit. This account is my account alone and not a joint account. Paypal contacted my wife and divulged all the inofromation about the account to her. They have violated my privacy by revealing all of the information to my wife. <br><br> Complaint ID (1774899) --- The associate said that my husbands credit will also be reported if I don't pay. This is for dental work I had done. My husband was my emergency contact. So I told him this is inaccurate as we both did not have work done and he said that if I don't pay that they will report my husbands credit. |
| **Topic 5** |
| Complaint ID (1302942) --- I have paid all copays and just rec 'd a bill from a collection agency. I have never rec 'd anything in the mail before on this. I called the Dr offfice and they say it was because visits that are being billed didnt have a referral but I always had a referral and if they told me that I would have gotten XXXX. I have attached proof that my primary care physician did send referrals. I feel as though this is some sort of way for them to bill people almost 5 years later knowing they will pay or it will kill their credit. I am in the XXXX and will be forced to pay something that is not accurate almost 5 years after the services. <br><br> Complaint ID (1291437) --- Paypal XXXX XXXX has continued to call my workplace and harass me AFTER I have specifically told them to DO NOT CALL MY EMPLOYEE ever again! Our receptionist is also a witness that they continue to call and this needs to stop. It is very disruptive to my job and against our company policy! |

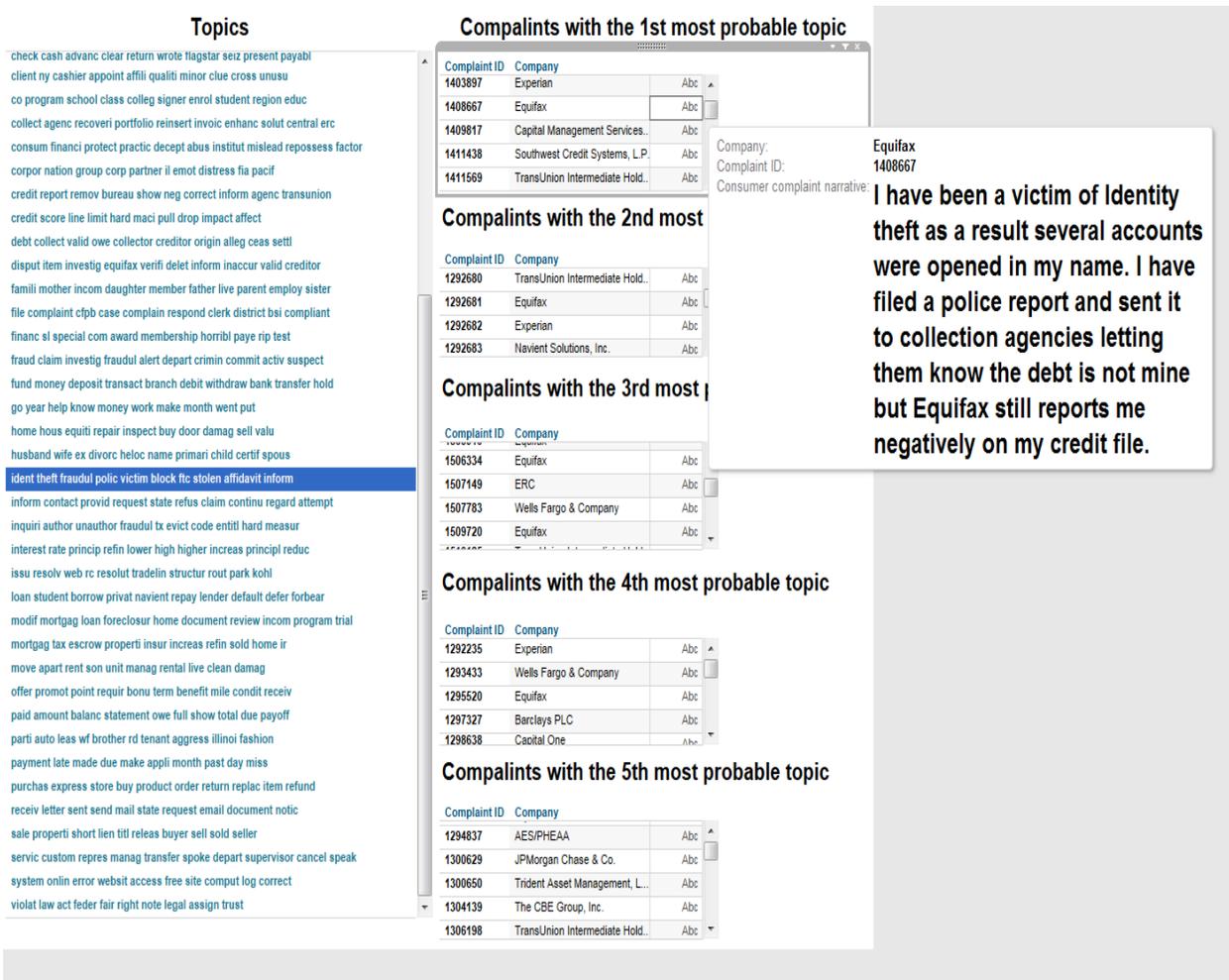

Figure 5. Overview of the Tableau viz on semantic similarities of the CFPB consumer complaints.

### 3.4 Topic Popularity Over Time

Another benefit of LDA analysis on the CFPB consumer complaints is that now each complaint is assigned with a mixture of topics (in our analysis top 5 topics with highest proportions are taken into account), then this would facilitate the analysis of the topics with respect to other fields of data in CFPB dataset. One of these fields is "Date Received" which represents the date CFPB received the consumer complaint. This field of data is particularly interesting in the sense that it would be possible to describe the frequency/popularity of the topics over time, and accordingly analyze the responsiveness of the financial companies with respect to these topics. For example, if it is observed that a topic trend is decreasing over time it will indicate that the CFPB regulations have taken into account the relevant topic, as well as the companies have been able to improve their customer service with respect to that topic.

First, it is required to define the topic popularity. Topic popularity represents the frequency of the topic appeared in the CFPB consumer complaints over time. However, as each complaint is assigned with a mixture of topics with different proportions, the frequency should be normalized (weighted) by the proportions of the topics in the complaints. For example assume two complaint narratives received in March 2015 were assigned with a mixture of topics {(3,0.35), (6,0.25), (17,0.2), (25,0.15), (33,0.05)}, and {(6,0.45), (17,0.3), (25,0.15), (33,0.08), (3,0.02)} respectively. The first element in ( , ) is the assigned topic ID, and the second element is its underlying proportion. Then the frequency of all the above topic IDs is two. However, the proportion does really matter instead of frequency, as it specifies how much of the complaints are explained by these topics. Hence, the topic proportions should be utilized to define the topic popularity.

Let $D_t = \{d_t^1, d_t^2, \ldots, d_t^{n_t}\}$ be the collection of documents (consumer complaints) received at time index $t$ (time unit is month in this analysis), $d_t^j$ be the $j$-th document in this collection ($\forall j = 1, \ldots, n_t$), where $n_t$ represents the total number of documents in $D_t$, then topic popularity for topic ID $i$ at time index $t$ is defined

$$Tp_{i,t} = \frac{\sum_{t=1}^{n_t} \theta_{i,d_t^j}}{n_t} \quad (\forall i = 1, \ldots, K; \forall t = 1, \ldots, T) \quad (3)$$

where $\theta_{i,d_t^j}$ denotes the proportion of topic ID $i$ assigned to document $d_t^j$, $K$ is the total number of topics (here is 40), and $T$ is the total number of time indices (will be discussed later). From Eq. (3), the nominator represents the frequencies of topic ID $i$ assigned to the documents received at time index $t$ (i.e. $D_t$) normalized by its proportions, and the denominator is used to further normalize the topic frequency metric over the number of documents received at time index $t$. The denominator normalization is critical in order to remove the bias effect of the number of documents received over time, and consequently represent the topic popularity metric in range of 0 and 1. Now the topic popularity for the above example in March 2015 (e.g. $t = 1$) is computed as $Tp_{3,1}$=0.185, $Tp_{6,1}$=0.35, $Tp_{17,1}$=0.25, $Tp_{25,1}$=0.15, and $Tp_{33,1}$=0.065.

Using Eq. (3) it would be possible to explore the CFPB topics popularity over time. Figure 6 shows the time trends of 12 selected topics from March 2015 to July 2016[7]; the chosen time unit is "month", hence $T = 17$ in Eq. (3) (the readers are referred to Appendix A to view the time trends for other CFPB topics). The time trends in Figure 6 are generally perceived as the popularity of their corresponding topics characterized with increasing, decreasing, or variable trends. The topics "Account Management", "Credit Reporting", "Rewards and Promotions", "Credit Score", and "CFPB" generally have increasing trend. Decreasing trend is evident for "Loan/Student Loan", "Harassment", and "Mortgage/Loan Modification and Foreclosure". Other topics including "Fraudulent/Unauthorized Inquiry", "Fund and Deposit", "Debt Collector", and "Dispute" have been highly variable, often accompanied with a high spike.

In Figure 6 "Credit Reporting", "Harassment", "Account Management", and "Debt Collector" have been the most popular topics (largest $Tp_{i,t}$'s ), respectively. This indicates that the consumer complaints have been mainly associated to these topics over time. Hence, the CFPB examiners might consider more efforts to address these topics as they contain the most proportions of the consumer complaints. Furthermore, some topics have been found with a sudden increase, decrease, or spike in their time trends. A high spike occurred in October 2015 in "Fraudulent/unauthorized Inquiry" and "Fund and Deposit" (about 100% increase in its computed topic popularity). In the time trend of "Rewards and Promotions", two consecutive high increases occurred in May 2016 (about 100% increase), followed by a much higher increase in June 2016 (about 400% increase). The topic popularity computed for "Mortgage/Loan Modification and Foreclosure" presents a high decrease in July 2016 (about 100% decrease). The sudden changes in the topics above might be relevant to some issues, policies, or new regulations that have caused their occurrences. Hence, a more thorough study is required to reveal the main causes of the sudden changes for these topics. Such discussion is provided in the next Section.

---

[7] CFPB started publishing online the consumer complaint narratives from March 2015; hence the timeline considered in this analysis is from March 2015 through July 2016.

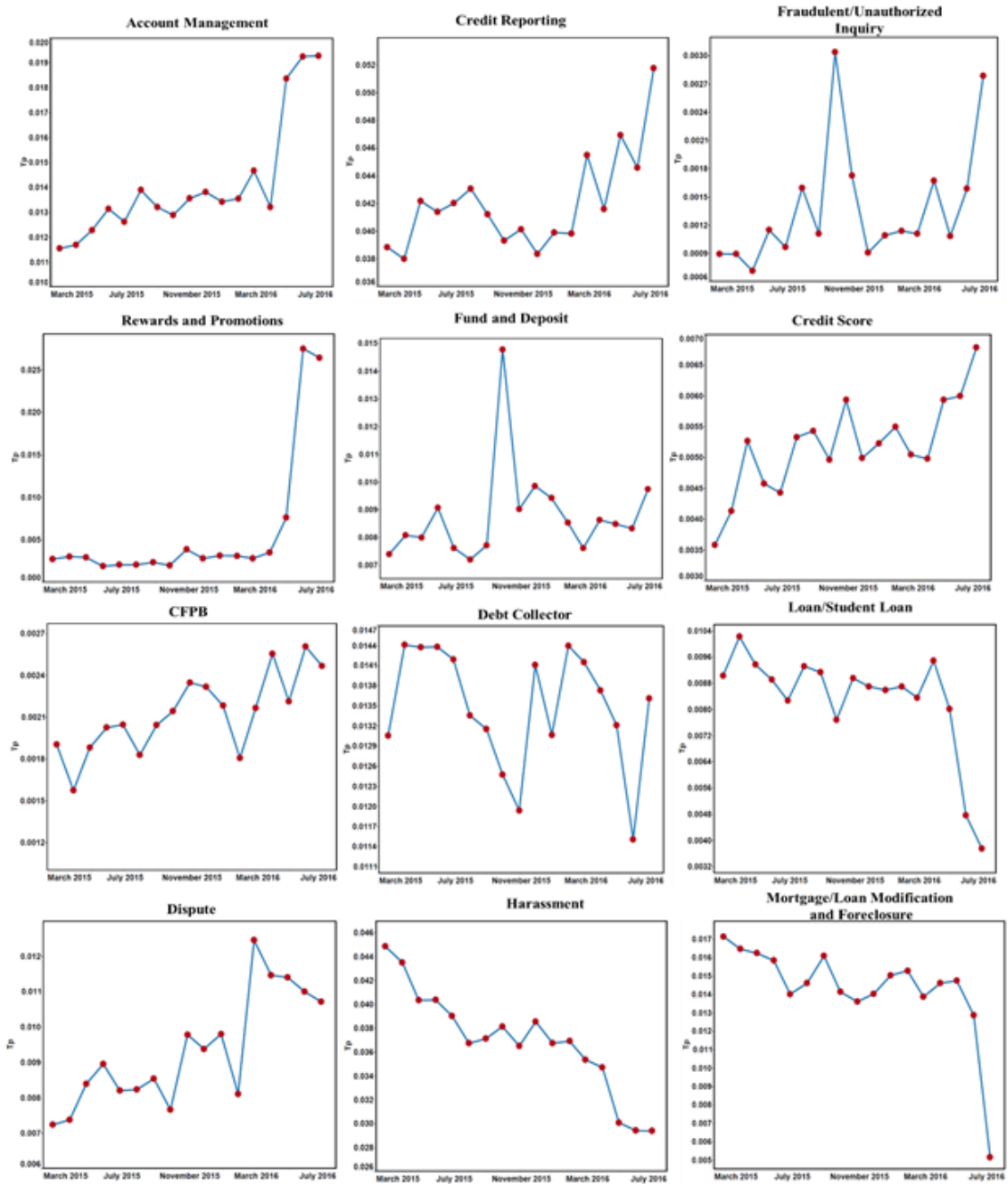

Figure 6. Time trends of 12 selected topics from March 2015 to July 2016.

## 4. Discussion

The results demonstrate the utility of topic modeling for the CFPB consumer complaints. The proposed approach using LDA created coherent and semantically meaningful topics/clusters from consumer complaints. These topics not only summarize a large collection of documents/complaints into human-interpretable decomposition of the texts, but also help practitioners discover new contents that might have been neglected by the CFPB. In the following it is discussed that the extracted topics are semantically cohesive, argued why and how the proposed approach could outperform the CFPB convention in labeling the complaints, and finally elaborated on the insights gained from topic popularity and time trend analysis using the proposed approach.

One of the fields in the CFPB datasets is "Issue" which describes the primary reason for the consumer complaints. These issues are predetermined labels defined by the CFPB experts, and provided in a drop-down menu for the consumers at the time of complaint submission. Once the consumers want to submit their complaints through the CFPB website, they should select their corresponding issue using this drop-down menu. A snapshot from the CFPB complaint submission portal is presented in Figure 7. There is a box in the top of this portal where consumer imports complaint narratives, and at the bottom of the portal a drop-down menu exists through which consumer choses the corresponding issue (e.g. Bankruptcy, Billing disputes, etc.) from the predetermined labels. This is the CFPB convention in labeling the consumer complaints.

In total there are 90 labels in the issue field in the CFPB dataset, however, more than 99% of the consumer complaint categorized by the top 60 labels with highest frequencies. A list of these 60 labels, followed by a plot presenting their normalized frequencies is provided in Appendix B. Reviewing these labels divulges their correspondences with the extracted topics which are also presented in Appendix B. In general there is one-to-one correspondence between the predetermined labels and the extracted topics; there are a few labels correspond to a combination of two or more topics though. Hence, this verifies the extracted topics are semantically cohesive as they can capture the CFPB practitioner's intuitions about the potential consumer complaint issues.

Moreover, there are a few topics that are not corresponded to any labels. For example there exists no one-to-one correspondence between the topics including "CFPB", "Family", and "Divorce and Ex-spouse", with the predetermined labels. This means there exist some new issues (labels) that might have been neglected by the CFPB, but have been captured using LDA. This is certainly one of the main benefits of the proposed approach as it would be able to identify new/evolving issues which is not possible by the CFPB labeling conventions.

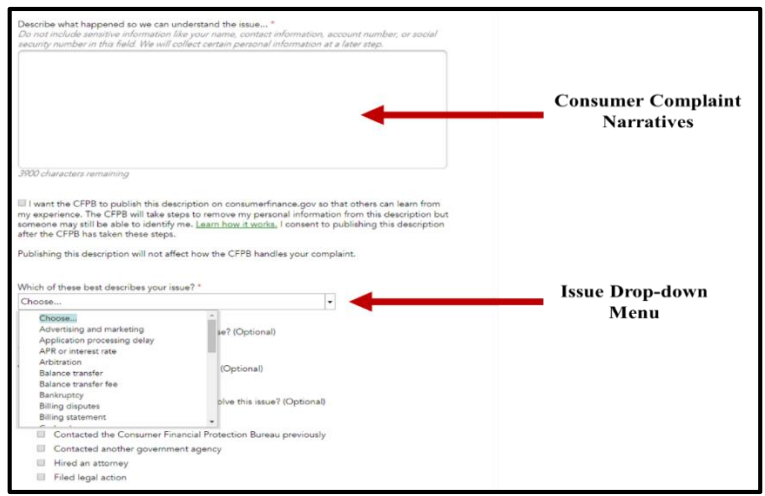

Figure 7. Snapshot from a consumer complaint portal on the CFPB website. The user would be able to select the corresponding issue for his/her complaint through a drop-down menu including 90 predetermined labels.

For example, the topic CFPB does not relate to any predetermined labels, but it has been extracted from the consumer complaints. This basically indicates that the consumers used relatively frequently the words relevant to the topic of CFPB in their complaint narratives. Figure 6 shows the topic popularity of CFPB which generally has an increasing trend over time. This might be due to increasing consumers' perception and expectation of CFPB. Filing a complaint against a company with the hope that the CFPB would pursue this case, and having not heard back from the CFPB or the company could be a main reason to file another complaint but this time against the CFPB itself. An example of such a narrative (complaint ID 1295147) is presented in the following:

"*I filed a complaint with cfpb against Ocwen and was led to believe that cfpb would pursue that company and make them repay monies they illegal took from an Escrow account. I was told by cfpb that I* *would have a response by XXXX/XXXX/15. It is now XXXX/XXXX/15 and I have not heard from either. I guess you guys just accept complaints and do nothing about them. Just another new, but not novel bureaucracy to inspire hope but deliver exasperation. Thanks a lot people for putting my taxpayer dollars to good use, once again. Case # XXXX XXXX*".

It should be emphasized that other statistical approaches are not capable of detecting the evolving topics in an automated way. For example meta-analysis has been used to conduct statistical inference from multiple studies with the aim of detecting relationships among evolving topics in a field of research (e.g. Song et al., 2018). However, it needs a manual review of all the studies which is a time consuming process.

Another benefit of the proposed approach is that it assigns a mixture of topics to a consumer complaint computed by posterior inferences using whole complaint collections. However, the CFPB conventional labeling approach is based on consumer input selected from the drop-down menu on the CFPB submission portal. There are two advantages with our proposed approach in this regard. First of all, the consumer selects the label best describing his/her complaint issue from the drop-down menu; hence due to the lack of knowledge or acquaintance of the consumers with the labels, they might select a wrong label (mislabeling), or if the issue they would like to complain about might not be relevant to any of the labels provided in the drop-down menu, then they will be forced to select a label that does not best describe their issue. Second of all, the consumer complaints might be about multiple issues (labels), but due to the limitation of the CFPB labeling convention they can only select one label.

These two shortcomings are addressed by our proposed approach through (1) topic assignment is carried out by posterior inferences, and (2) a mixture of topics is assigned to each consumer complaints. Let us consider the complaint narrative with complaint ID 1313544 presented in Figure 3. Our proposed approach assigns a mixture of topics (top 5 topics) including "Auto Loan/Dealership", "Collection Agency", "Communication", "Credit Reporting", and "Customer Service". However, the conventional CFPB labeling approach labels this complaint as "Cont'd attempts collect debt not owed". By reviewing the complaint narratives it can be realized that our proposed approach is more reliable in best describing the issue rather than the CFPB labeling convention.

Another benefit of the proposed approach is its topic popularity over time analysis. The time trend of topic popularity can be certainly used as a quantitative metric to evaluate the quality and effectiveness of the CFPB regulations in creating a consumer oriented culture in decision making processes of financial institutions. From Figure 6, "Harassment", "Loan/Student Loan", and "Mortgage/Loan Modification and Foreclosure" generally have decreasing trends (especially within the last 4 months). For instance, the "Harassment" topic essentially refers to consumer complaints about debt collectors' actions and communication tactics; consumers most commonly complain that the collectors harass them and request amounts that consumers do not owe. To protect the consumers from harassment issues, federal and state officials consider the debt collection as the first priority, and accordingly since October 2012 the CFPB started, as the first federal agency, using its enforcement authority to routinely supervise debt collectors[8]. Hence, the routine supervisory authority of the Bureau over the debt collectors must have resulted in a decreasing trend in the "Harassment" topic popularity over time. Another example is the decreasing trend of the "Mortgage/ Loan Modification and Foreclosure" topic which could be mainly caused by the mortgage servicing rule became effective by the CFPB in 2014. This rule was issued with the aim of expanding foreclosure protections for struggling borrowers and homeowners, and it has been consistently updated since then[9].

As opposed to decreasing trend, increasing trend or high variability of topic popularity could indicate that the CFPB regulations have not been adequately effective in addressing the consumer complaints with respect to those topics. For example, in spite of a new Regulation (V) (Fair Credit Reporting) being established by the CFPB in November 2012 (Mogilnicki and Malpass, 2012) to enhance the standards for communication and use of information bearing on a consumer's creditworthiness, and credit standing, "Credit Reporting" has been one of the top three popular topics over the past few years with generally an increasing trend over time (see Figure 6). "Account Management" is another topic with an increasing trend where a high increase occurred in May 2016. Very recently in September 2016, the CFPB fined Wells Fargo $100 Million for secretly opening unauthorized accounts, and shifting funds from consumers' existing accounts into these new accounts without their permission (it dates back to at least five years). Hence, to protect the consumers from the similar issues, the CFPB probably needs to devote further investigations to identify the reason behind this high increase in the "Account Management" topic popularity. Similarly the "Rewards and Promotions" topic popularity increased significantly in May and June 2016; this might be of interest to the CFPB for further investigations as well.

---

[8] Defining Larger Participants of the Consumer Debt Collection Market, 77 FR 65775 (Oct. 31, 2012), 12 CFR1090.
[9] http://files.consumerfinance.gov/f/201301 cfpb servicing-rules summary.pdf

Although the effectiveness of the CFPB regulations could be the main reason for the overall trend of topic popularity, the financial institutions' role should never be neglected as they are responsible to employ the CFPB regulations as part of their organizational decision making processes. Hence, the level of their commitment and compliance to these regulations could be another reason for decreasing or increasing trends of topic popularity over time. One of the fields in the CFPB dataset is "Company" which indicates to which financial institution the received consumer complaint relates. In the future, it would be of interest to monitor the topic popularity at the level of each financial institution. This would certainly provide remarkable insights into the community, and enables us to investigate on which institutions have better compliance to the CFPB regulations according to the topic popularity computed using their consumer complaints.

As an early follow-up investigation to our future research, we refer the readers to topic popularity of "Fund and Deposit" which can be seen in Figure 8. The figure shows the time trend of this topic for all the financial companies reported in the CFPB dataset. Due to the space limit, the captions of only a very few of these companies are illustrated (i.e. Wells Fargo, Bank of America, Citibank, and Empowerment Ventures, LLC). Our analysis discloses a high spike in October 2015, which is primarily associated with Empowerment Ventures, LLC (RushCard). RushCard is a financial company providing prepaid cards for unbanked consumers. In October 2015, thousands of RushCard customers experienced problems as a result of a software conversion. Their consumers were frustrated because they were not able to access their cash, check their balances, or deposit any money into their accounts[10]. As such, they started sending out complaints to the CFPB. An analysis of the complaint narratives, similar to that which is discussed in this document, may have led to detection more quickly.

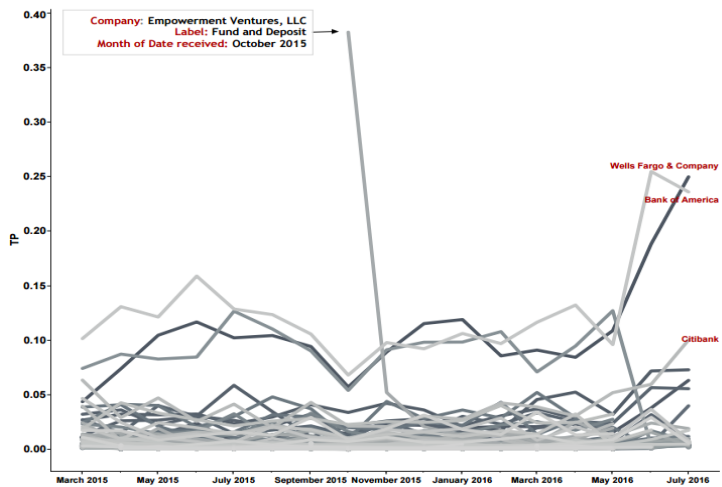

Figure 8. Topic popularity of Fund and Deposit given all the financial companies in the CFPB dataset.

## 5. Conclusion

A decision support system (DSS) based on LDA was proposed in this paper to reveal useful information from the CFPB consumer complaint narratives. LDA generates semantically meaningful topics/clusters to summarize the consumer complaints into a mixture of topics that would not be possible by human annotations as reading large volumes of narrative data and interpreting them is very time consuming and difficult. The topic popularity/time trend analysis of these topics can be useful in evaluating the effectiveness of the CFPB regulations in addressing the consumer complaint issues. It was demonstrated that the CFPB regulations have been successfully enforced for "Harassment", "Mortgage/Loan Modification and Foreclosure" and "Loan/Student Loan" as they were the topics with decreasing time trend, and for the topics including "Credit Reporting", "Credit Score", "Account Management",

---

[10] http://money.cnn.com/2015/10/15/news/companies/rushcard-russell-simmons/

"Fraudulent/Unauthorized Inquiry" and "Dispute" the regulations have not been adequately effective due to their increasing or highly variable trends.

The proposed approach can be certainly useful in monitoring the consumer complaint narratives for emerging topics. It can be easily operationalized (in a number of steps including data preprocessing, LDA based topic modeling, and topic popularity analysis) and run monthly (even daily) to automate the prediction of topic assignments of the new complaint narratives that will be received in the future. Similar to the procedures explained in this paper, monthly (daily) topic popularity analysis of the new complaints can reveal emerging topics (e.g. as "Rewards and Promotions" and "Account Management" were revealed in this paper) based on high increases captured in their very recent time trend. The proposed approach is never intended to replace human analysts, instead it is designed as a decision support tool for the existing CFPB analysts to investigate consumer complaints more efficiently and effectively, and eventually improve consumer protections from unfair, deceptive or abusive practices in the financial markets.

# Appendix A

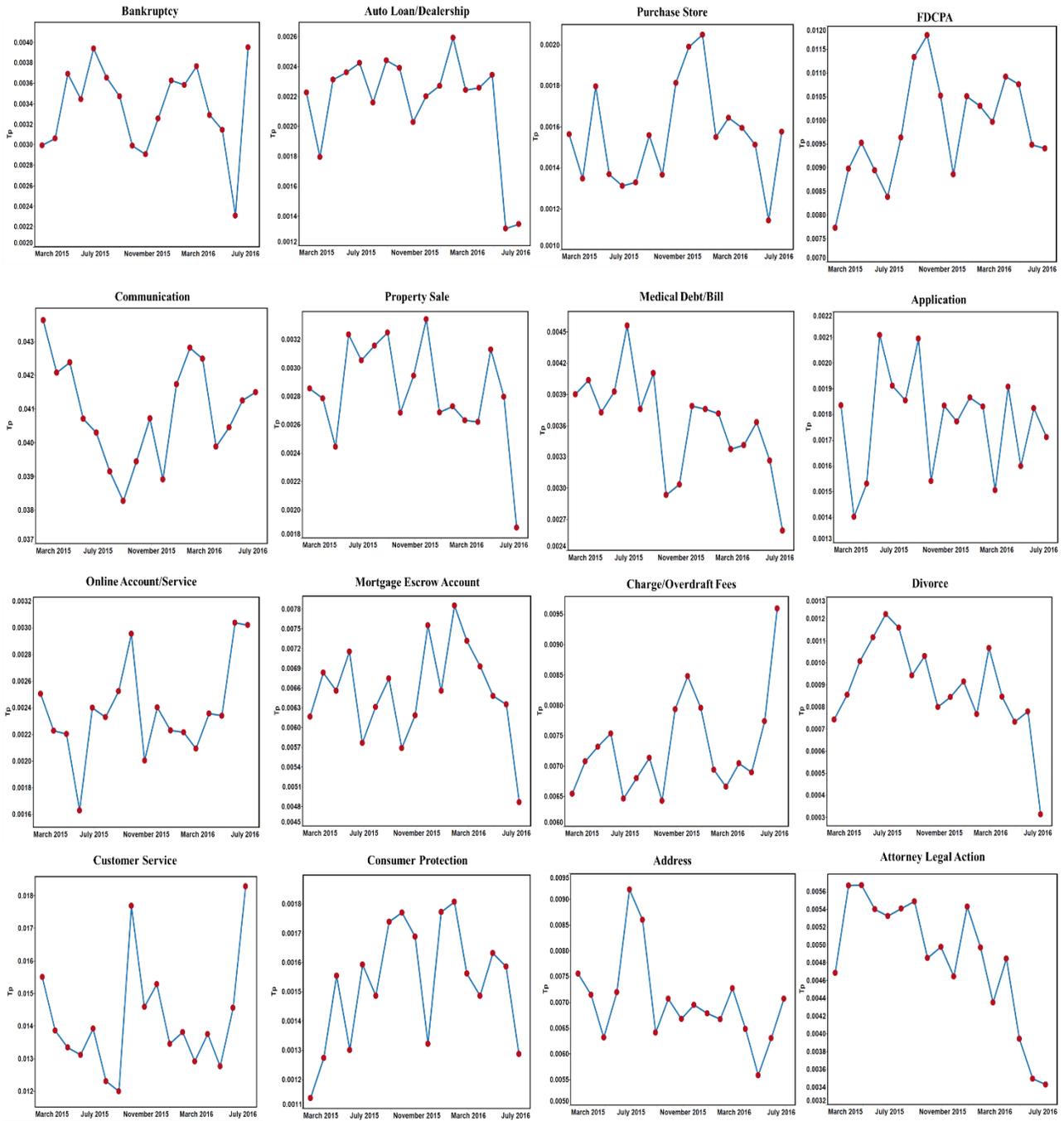

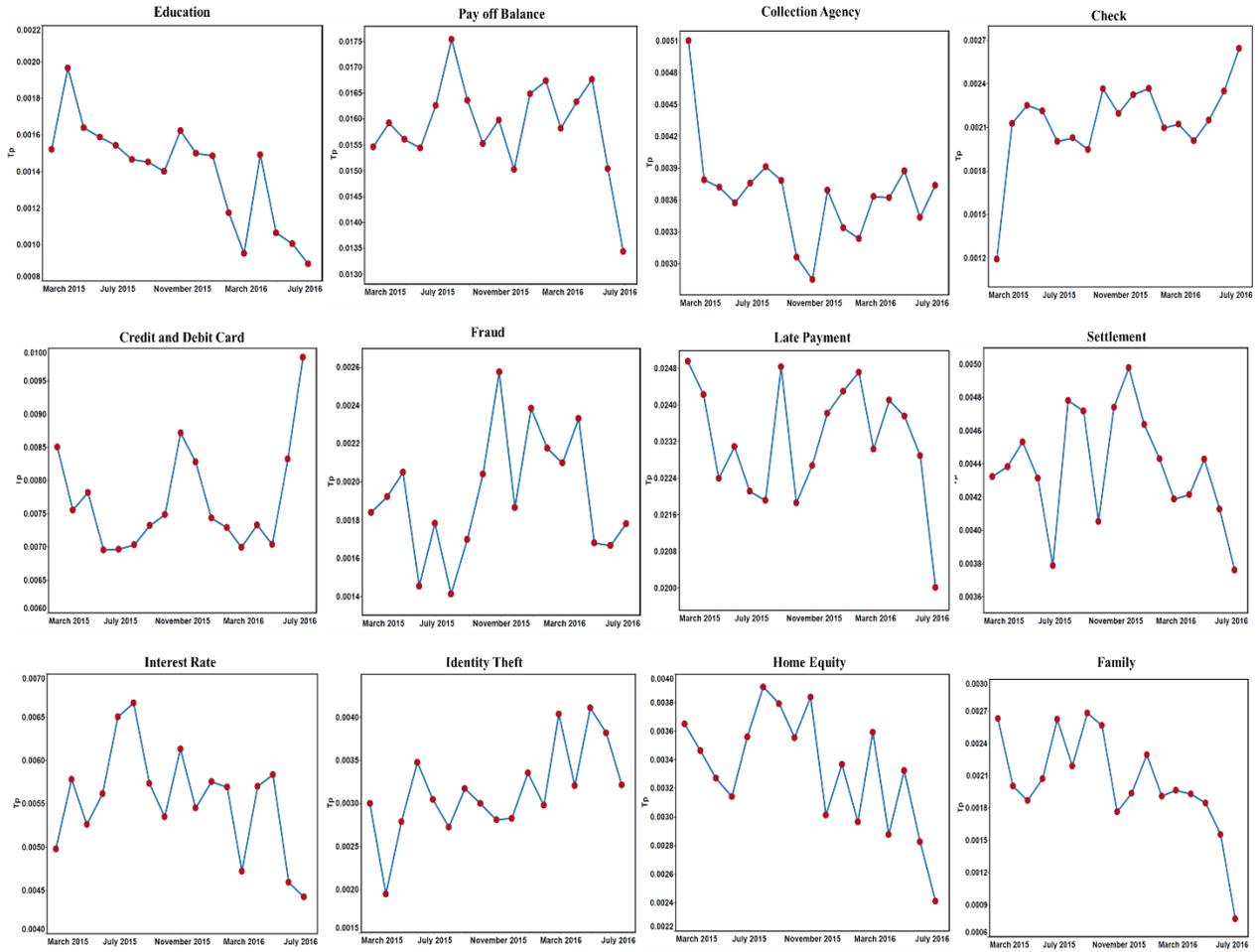

Figure A.1. Time trend analysis of the rest of topics not included in Figure 6.

**Appendix B**

Table B.1. Top 60 labels of the issue field in the CFPB dataset, and their correspondence with topics extracted using LDA

| Label | Topic ID | Label | Topic ID |
|---|---|---|---|
| Incorrect information on credit report | 1 | Fraud or scam | 32 |
| Cont'd attempts collect debt not owed | 9 | Customer service / Customer relations | 4 |
| Loan servicing, payments, escrow account | 2,28 | Delinquent account | 0 |
| Loan modification, collection, foreclosure | 2,20 | Advertising and marketing | 36,37 |
| Disclosure verification of debt | 6,36 | APR or interest rate | 26 |
| Communication tactics | 9 | Charged fees or interest I didn't expect | 33 |
| Account opening, closing, or management | 12 | Late fee | 0,33 |
| Credit reporting company's investigation | 10 | Rewards | 36 |
| Managing the loan or lease | 2,28 | Shopping for a loan or lease | 2,3,15,34 |
| False statements or representation | 1,23 | Billing statement | 1,21, 47 |
| Application, originator, mortgage broker | 39 | Unauthorized transactions/trans. issues | 17,33 |
| Taking/threatening an illegal action | 9 | Credit card protection / Debt protection | 27 |
| Dealing with my lender or servicer | 2,28 | Transaction issue | 0,13, 16,33 |
| Deposits and withdrawals | 35 | Payoff process | 38 |
| Improper contact or sharing of info | 1,23 | Credit determination | 27 |
| Billing disputes | 16 | Managing, opening, or closing account | 12 |
| Unable to get credit report/credit score | 23,27 | Other fee | - |
| Problems when you are unable to pay | 0,38 | Credit line increase/decrease | 27 |
| Settlement process and costs | 24 | Other transaction issues | - |
| Problems caused by my funds being low | 35 | Unsolicited issuance of credit card | 27 |
| Other | - | Can't contact lender | 2,20 |
| Identity theft / Fraud / Embezzlement | 19 | Balance transfer | 35,38 |
| Can't repay my loan | 2, 20 | Money was not available when promised | 33,35 |
| Improper use of my credit report | 17 | Received a loan I didn't apply for | 2,20 |
| Taking out the loan or lease | 2, 20 | Can't stop charges to bank account | 33 |
| Closing/Cancelling account | 12 | Getting a loan | 2,20 |
| Using a debit or ATM card | 27 | Payment to acct not credited | 0,38 |
| Credit decision / Underwriting | 23,27 | Other service issues | - |
| Making/receiving payments, sending money | 0,37, 38 | Adding money | 33, 35 |
| Credit monitoring or identity protection | 14,27 | Bankruptcy | 25 |

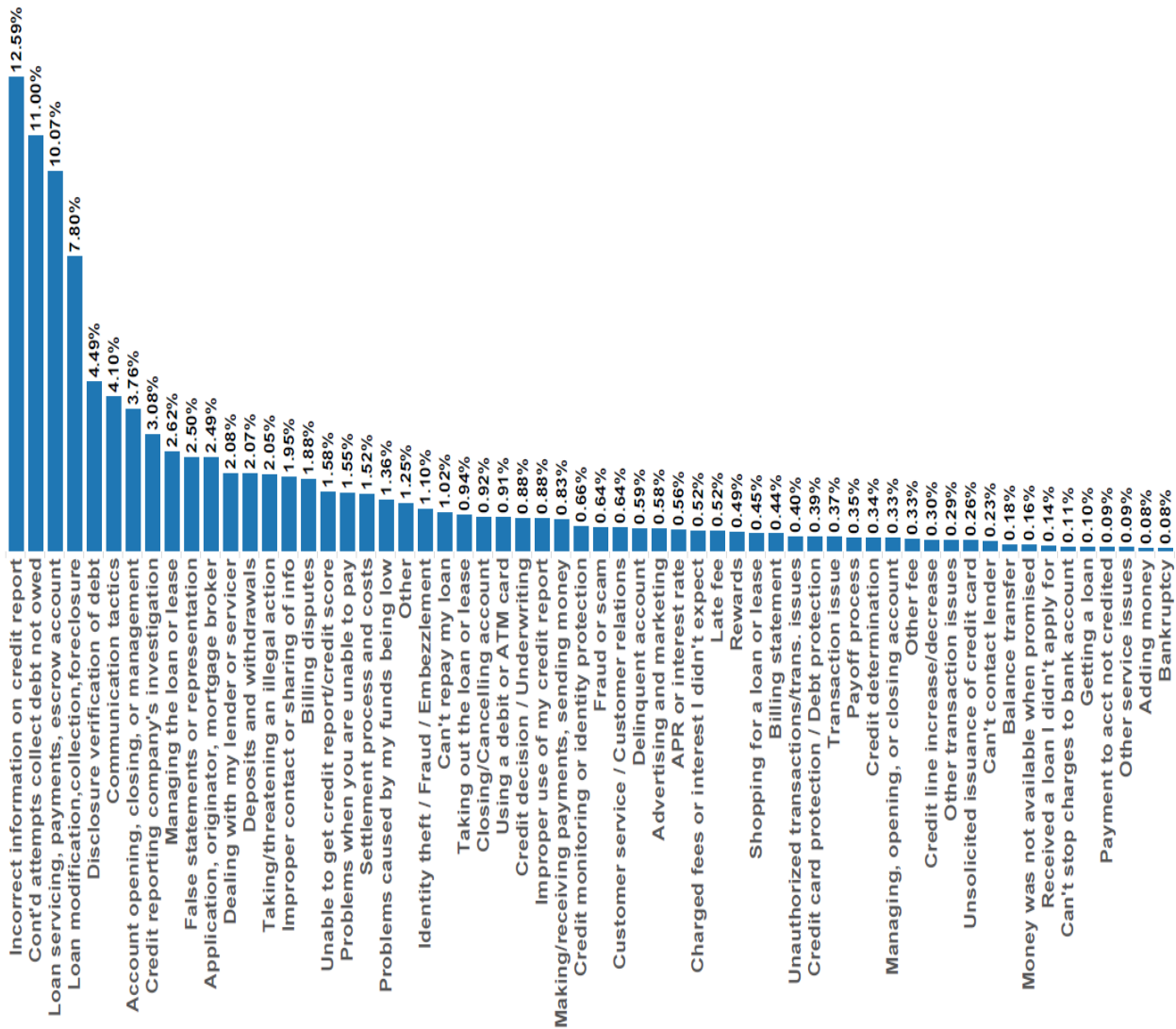

Figure B.1. Normalized frequency of top 60 issues in the CFPB dataset. These issues cover labeling of more than 99% of the whole consumer complaints in the dataset.